\documentclass{aa}
\usepackage{amsmath}
\usepackage{graphicx}
\usepackage{natbib,twoopt}
\usepackage[breaklinks=false]{hyperref} 
\usepackage[varg]{txfonts}
\usepackage{ulem}
\usepackage{xcolor}
\usepackage{nicefrac}

\bibpunct{(}{)}{;}{a}{}{,}

 \newcommandtwoopt{\citeads}[3][][]{\href{http://adsabs.harvard.edu/abs/#3}%
                                        {\citealp[#1][#2]{#3}}}
 \newcommandtwoopt{\citepads}[3][][]{\href{http://adsabs.harvard.edu/abs/#3}%
                                        {\citep[#1][#2]{#3}}}
 \newcommandtwoopt{\citetads}[3][][]{\href{http://adsabs.harvard.edu/abs/#3}%
                                        {\citet[#1][#2]{#3}}}
 \newcommandtwoopt{\citealtads}[3][][]{\href{http://adsabs.harvard.edu/abs/#3}%
                                        {\citealt[#1][#2]{#3}}}

\begin{document}

\title{Structures in circumbinary disks: Prospects for observability}
\titlerunning{Structures in circumbinary disks:  Prospects for observability}

\author{Jan Philipp Ruge\inst{1,}\thanks{ruge@astrophysik.uni-kiel.de} \and Sebastian Wolf\inst{1} \and Tatiana Demidova\inst{2} \and Vladimir Grinin\inst{2,3}}

\institute{Institute of Theoretical Physik and Astrophysics, Kiel University, Leibnitzstr. 15, 24098 Kiel, Germany \and 
Pulkovo Astronomical Observatory, Russian Academy of Sciences, Pulkovskoe shosse 65, St. Petersburg, 196140 Russia \and 
Sobolev Astronomical Institute, St. Petersburg State University, Universitetskii pr. 28, St. Petersburg, 198504 Russia}

\authorrunning{Ruge et al.}
\date{Received 25. April 2013/ Accepted 06. May 2015}

\abstract{During the past decade circumbinary disks have been discovered around various young binary stars. Hydrodynamical calculations indicate that the gravitational interaction between the central binary star and the surrounding disk results in global perturbations of the disk density profile.}
{We study the observability of characteristic large-scale disk structures resulting from the binary-disk interaction in the case of close binary systems.}
{We derived the structure of circumbinary disks from smoothed-particle hydrodynamic simulations.
Subsequently, we performed radiative transfer simulations to obtain scattered light and thermal reemission maps.
We investigated the influence of the binary mass ratio, 
the inclination of the binary orbit relative to the disk midplane, 
and the eccentricity of the binary orbit on observational quantities.}
{We find that ALMA will allow tracing asymmetries of the inner edge of the disk and potentially resolving spiral arms if the disk is seen face-on. 
For an edge-on orientation, ALMA will allow detecting perturbations in the disk density distribution through asymmetries in the radial brightness profile.
Through the asymmetric structure of the disks, areas are formed with a temperature $2.6$ times higher than at the same location in equivalent unperturbed disks.
The time-dependent appearance of the density waves and spiral arms in the disk affects the total re-emission flux of the object by a few percent.
}{}
\maketitle

\section{Introduction}
During the past three decades the formation of single stars was explored in detail \citepads[e.g., see the review][]{2007ARA&A..45..565M}. 
Circumstellar disks, which are formed by the conservation of angular momentum during the molecular cloud collaps, play a keyrole in this process 
\citepads[e.g.,][]{2005ApJ...631.1134A,2006ApJ...653.1480W,2007ApJ...671.1800A,2009ApJ...701..260I,2009ApJ...700.1502A,2010ApJ...723.1241A}. 
The lifetime of these disks, which have been observed from the optical to the millimeter wavelength range, is limited to a few million years \citepads{2000prpl.conf..559N,2000prpl.conf..703M,2001ApJ...553L.153H,2012A&ARv..20...52W}. 
In the optical and near-infrared regime, the scattering of stellar radiation on the surface dominates the appearance of these disks, while the thermal reemission is stronger at far-infrared to millimeter wavelengths. 
High-angular observations across this wavelength range are important for the analysis of the disk structure and composition as well as for their evolution \citepads[e.g.,][]{2011A&A...527A..27S, Graefe13}.
 \par
However, since the mid-1990s it became evident that a large portion of stars forms as binaries \citepads[e.g.,][]{2009ApJ...704..531K, 1994MNRAS.271..999B, 1999A&A...341..547D}. Meanwhile, gas and dust disks have been discovered in several young binary systems \citepads[e.g.,][]{1999A&A...348..570G,2002ApJ...575..974M,2007A&A...476..229D,2012ApJ...749...79R}. 
They come in two flavors: 
First, the systems where at least one component of the binary hosts a circumstellar disks, as studied by \citetads{2011IAUS..276...50R,2011A&A...528A..93R};
and second, binaries with a circumbinary disk (e.g., GG tau \citealtads{2011A&A...528A..81P}; V4046 Sgr., \citealtads{2010ApJ...720.1684R}).
In the second case (circumbinary disks), the orbital motion of the central binary is predicted to induce periodic perturbations, spiral density waves, and gas flows in the disk \citepads{1996ApJ...467L..77A,1997MNRAS.285..288L,2002A&A...387..550G,2004A&A...423..559G,2007AstL...33..594S,2010ApJ...708..485H,2010ARep...54.1078K,2012ApJ...757L..29A,2011MNRAS.413.2679D,2012MNRAS.427.2660K}. 
Characteristic features resulting from this interaction are inner cavities with the radius of about twice the semi-major axis of the binary \citepads[e.g.,][]{1994ApJ...421..651A,2002A&A...387..550G}. 
This binary-induced inner cavity results in a flux-deficit in the mid-infrared wavelength range and a large inner hole in the brightness distribution of these objects (at least at sub-to-millimeter wavelengths). This appearance is similar to what has been prediced and found in the case of transitional disks 
\citepads[e.g.,][]{1994ApJ...421..651A,2009arXiv0901.1691E,2010ApJ...712..925C}.
The existence of circumbinary planets was shown through the discovery of Kepler 34b \citepads{2012Natur.481..475W}, for instance. However, the effect of the binary-induced perturbations on the process of planet formation in circumbinary disks is still under debate \citepads[e.g.,][]{2012ApJ...754L..16P,2014ApJ...782L..11L,2014arXiv1406.1357T}. Observational constraints that allow verifying the physical processes of the predicted binary-disk interactions are required.
For this purpose, we investigate the feasibility of observing characteristic binary-induced perturbations in circumbinary disks. In particular, we focus on the case of embedded close binary systems. Previous studies have shown that the orbital motion of binary-induced large-scale density perturbations in the disks can cause a large-amplitude brightness modulation of the system (\citealtads{2007AstL...33..594S}; \citealtads{2010AstL...36..422D}; \citealtads{2010AstL...36..498D}; \citealtads{2010AstL...36..808G}). In particular, the perturbation of the inner part of the disk affects the illumination of its outer regions, resulting in an asymmetric image of the disk if observed face-on \citepads{demidova}.

Based on selected circumbinary disk models resulting from 3D smoothed-particle hydrodynamic (SPH) simulations, we perform radiative transfer calculations to derive the thermal structure and observable quantities of these circumbinary disks. To trace the densest regions in the disks best -- and thus the most prominent density perturbations -- one has to make sure to observe in a wavelength range where they are optically thin. Combined with the second requirement, namely to spatially resolve structures on the sub-100AU scale, we expect the Atacama Large Millimetre/submillimeter Array (ALMA) to be the potentially best observatory for future studies. For this reason we prepared and analyzed simulated observations of circumbinary disks with this interferometer.\par

The paper is divided into three parts. First, we describe the SPH simulations in Sect. \ref{sec:sph} and the radiative transfer calculations in Sect. \ref{sec:radia}. All assumptions and the parameter setup are addressed in these sections. Predicted ALMA observations are presented in Sect. \ref{sec:ALMA}.

\section{Circumbinary disk structure}
\label{sec:sph}

  \begin{figure}[t]
      \resizebox{\hsize}{!}{\includegraphics{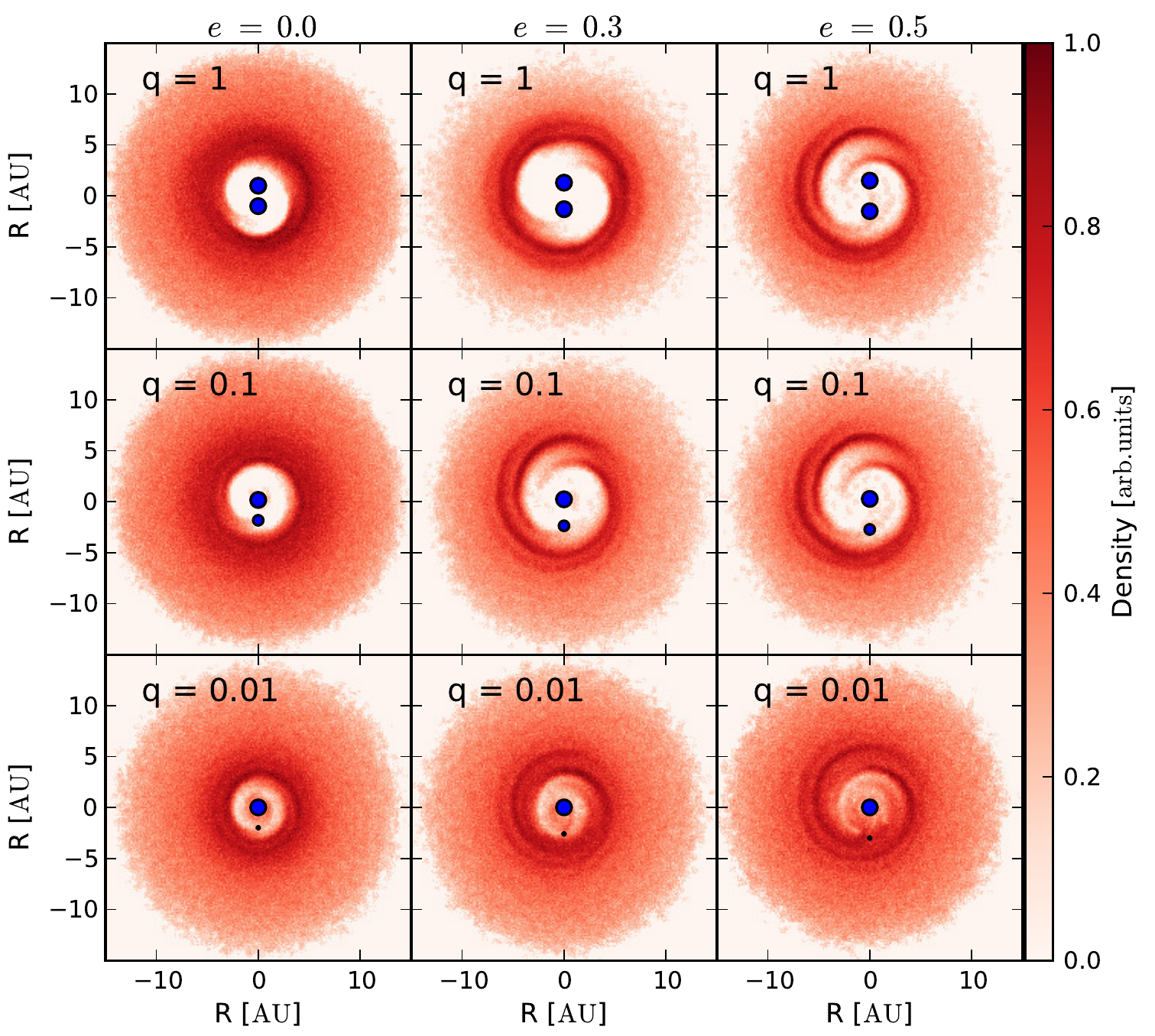}}
  \caption{Illustration of the projected disk density distribution in a face-on orientation for models with an eccentric binary orbit. The binary mass ratio $q$ is shown in the upper left corner of every subplot. The filled circles indicate the positions of the binary stars.}
  \label{fig:density_ec}
  \end{figure}

As the basis of our study, the periodically changing gravitational interaction between the embedded binary system and the surrounding disk was calculated with 3D SPH simulations. In the following, we explain the underlying assumptions and parameter setup of our study. 
We applied the SPH code developed by \citetads{1996Ap.....39..141S}; for a more detailed description of the SPH models see \citetads{2010AstL...36..498D} and \citetads{2010AstL...36..422D}. 
For the follow-up radiative transfer calculations (see Sect. \ref{sec:radia}) we assumed a dust-to-gas mass ratio of $1:100$.\par

We focus on a simple disk setup and selected orbital elements of the binary and the binary mass ratio.\\
{\em Central binary:}
We assumed a close binary with a semi-major axis of $a=2.0 \,  \rm AU$, which corresponds to a period of $1.7\, \rm yr$ ($M_\text{prim} = 2.5\, \rm M_\odot$).\\
{\em Circumbinary disk:}
We assumed a fully isothermal disk. 
The effective viscosity of the disk \citepads{1973A&A....24..337S}, was set to $\alpha_s = 0.03$ (for details of implementating the viscosity in our simulations see \citetads{2007AstL...33..594S}. 
We assumed an initial pressure scale height of $h = \nicefrac{H}{r} = 0.1$. 
This value corresponds to a disk temperature of $T = 100\, \rm K$. 
We note, however, that the simulation of ALMA observations (see Sect. \ref{sec:ALMA-res}) was based on a 3D temperature distribution that was derived from 3D radiative transfer simulations on the basis of the density distribution resulting from the SPH simulations (see Sect. \ref{sec:sph}).
\begin{figure}[t]
      \resizebox{\hsize}{!}{\includegraphics{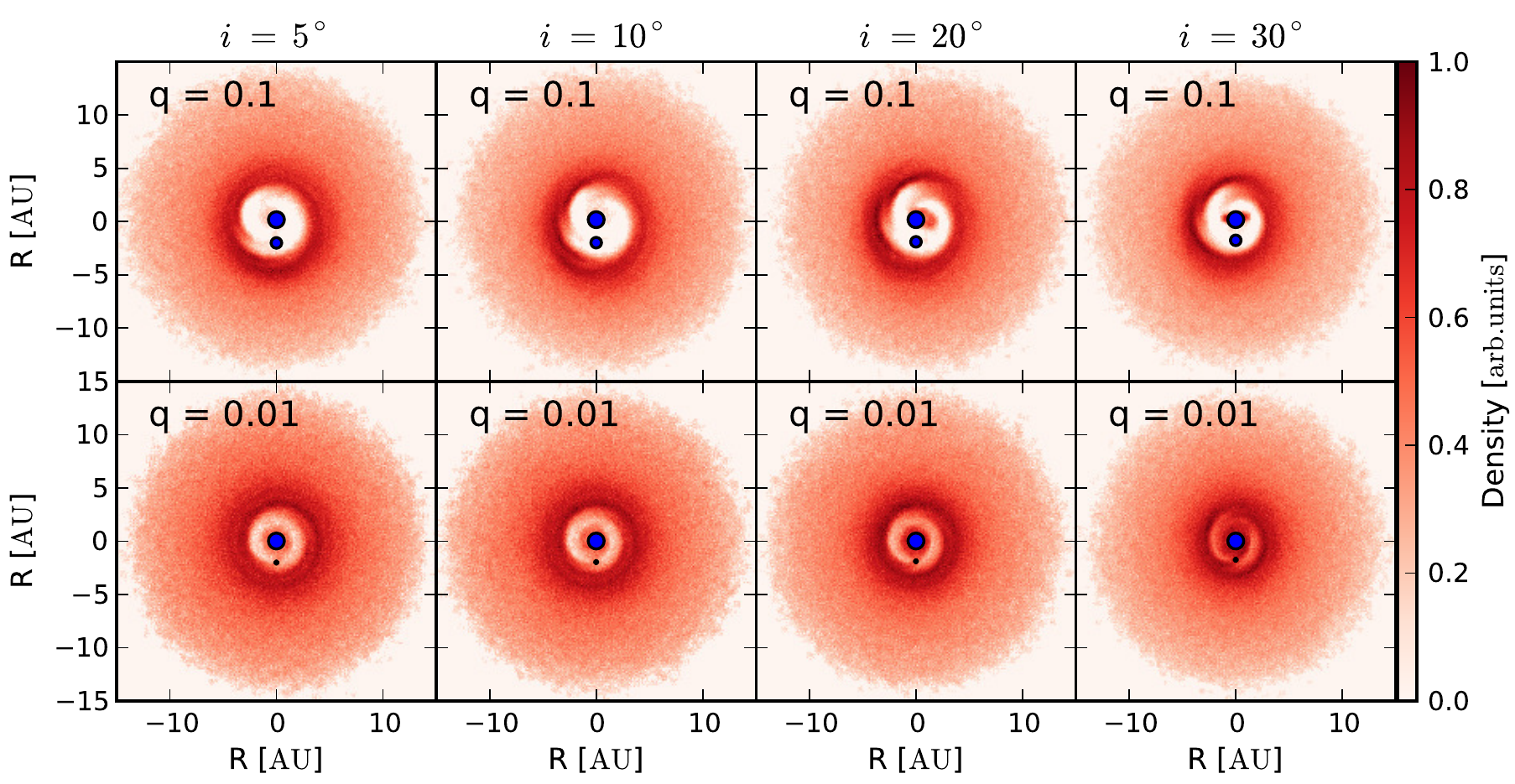}}
  \caption{Illustration of the projected disk density distribution in a face-on orientation for models with a binary orbit that is inclined relative to the initial disk midplane. The binary mass ratio $q$ is shown in the upper left corner of every subplot. The filled circles indicate the positions of the binary stars.}
  \label{fig:density_inc}
  \end{figure}
The calculations were performed using  $5 \times 10^5$ SPH particles.
For the radiative transfer post-processing, we selected the distribution of SPH particles after 30 orbits. This is a valid compromise that reflects the evolution of the density distribution in the inner part of a circumbinary disk sufficiently well. On one hand, the accretion rate onto the binary components saturates after 20 orbits. On the other, the system has lost $5\%$ of all particles as a result of accretion after $40$ orbits. The outer radius of the SPH simulated disks was set to $r_\text{out} = 15\, \rm AU$.
The total dust mass of the disk amounts to $M_\text{dust} = 4.4 \times 10^{-5} \, \rm M_\odot$.
For investigations that have to respect the influence of the a complete disk, we assumed an outer disk following the expression in Eq. \ref{glg:shakura} \citepads{1987ApJ...323..714K,1973A&A....24..337S}, which was added to the numerically calculated inner disk. This approach is necessary to take into account possible shadowing effects in close to edge-on orientations of the disk,
\begin{equation}
  \rho_\text{out}\left(r,z\right) \propto \left(\cfrac{R_{0}}{r}\right)^{\gamma} \exp\left\{-\left[ \cfrac{z}{h\left(r\right)} \right]^2 \right\}, \label{glg:shakura}
\end{equation}
where
\begin{equation}
 h\left(r\right) = h_0 \left(\cfrac{r}{R_{0}}\right)^{\beta}\text{.} \label{glg:hr}
\end{equation}
In Eq. \ref{glg:shakura} was $\gamma = 1$, $\beta = 5/4$, and $R_0 = 6a$. The parameter $h_0$ was chosen to fit the scale height of the numerically simulated disk at the radius $R_0$ (approximately $h_0=0.6a$). The transit between the numerically calculated density profile and the outer analytically described region was smoothed in a range of $2a$ with a Gaussian to avoid high density gradients.
Now, the outer radius was set to $R_\text{out} = 100\, \rm AU$ and the total dust mass amounts to $M_\text{dust} =  10^{-4} \, \rm M_\odot$ or $M_\text{dust} =  10^{-5} \, \rm M_\odot$. This leads to a mass ratio between the disk and primary component of the binary of $0.4\%$, hence effects of self-gravitation can be neglected \citepads{1994ApJ...421..651A}.

{\em Parameter space:}
We considered various values for the binary mass ratio $q = \nicefrac{M_\text{sec}}{M_\text{prim}}$, the eccentricity of the binary orbit, and its inclination to the midplane of the initial disk.
For binary mass ratios of $q = 1, \, 0.1$ and $0.01$, we explored binary orbits with eccentricities of $e = 0.0,\, 0.3$, and $0.5$. The orbits are coplanar to the midplane of the initial disk. Simulations of ALMA observations were performed when the secondary component is in apastron.
Figure \ref{fig:density_ec} illustrates the projected density distributions of these disks in a face-on orientation.

We also investigated binaries with a low-mass secondary component ($q = 0.1$ and $0.01$) on circular orbits that are inclined by $i = 5^{\circ},\, 10^{\circ},\, 20^{\circ}$, or $30^{\circ}$ to the midplane of the initial disk. In these cases, we post-processed the SPH simulation output with the secondary component perpendicular to the line of nodes and underneath the circumbinary disk plane. Figure \ref{fig:density_inc} illustrates the projected density distributions of these disks in a face-on orientation.
In addition to that, we calculated 21 snapshots of a disk model with $q= 0.5$, $i= 0^\circ$ and $e = 0.3$ to investigate structural changes in the disk structure during one period of the binary.\\
We note that once the binary clears the inner cavity and induces spiral arms (e.g., $q > 0.01$) it is not feasible to trace from the density distribution back to the binary mass ratio.

\section{Thermal disk structure and the observable quantities of the disks}
\label{sec:radia}

\begin{figure}[t]
    \resizebox{\hsize}{!}{\includegraphics{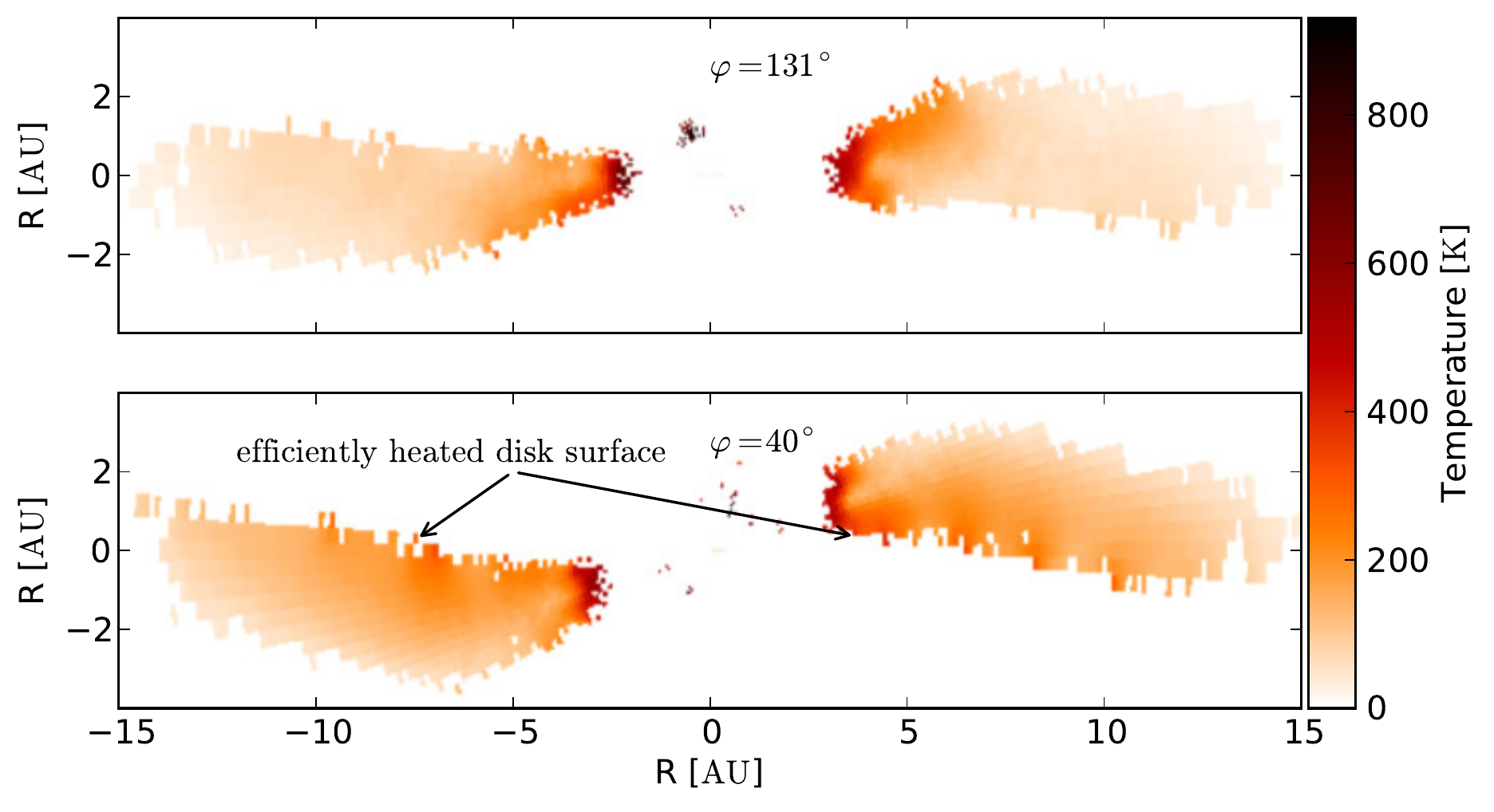}}
\caption{Temperature maps along vertical cuts through the density distributions of the disk model with the parameters $q= 0.1$, $i= 30^\circ$, and $e = 0.0$ shown in the top right corner in Fig. \ref{fig:density_inc}.}
\label{fig:vertical_disk_temp}
\end{figure}
\paragraph{Radiative transfer post-processing:}
To derive more realistic temperature distributions as the basis of simulated ALMA observations (see Sect. \ref{sec:ALMA-res}), 
the heating of the disk by the binary radiation was simulated using the Monte Carlo 3D continuum radiative transfer code MC3D 
(\citealtads{1999A&A...349..839W}; \citealtads{2003CoPhC.150...99W}). 
Subsequently, reemission maps were simulated. Only, for edge-on orientation of the disk, we used models with an analytically added outer disk (see Sect. \, \ref{sec:sph}).

{\em Dust density distribution, dust properties:}
We used a spherical grid that was subdivided linearly into azimuthal (angular resolution: $1.9^\circ$) and polar direction (angular resolution: $3.6^\circ$). In the radial direction the resolution is $0.133\, \rm AU$. As a result of the geometry of the circumbinary disks, the SPH particles are concentrated in $8\, \%$ of the grid close to the midplane of the initial disk. We substituted the distribution of SPH particles by a distribution of spherical dust grains that consisted of $62.5\, \%$ silicate and $37.5\, \%$ graphite (optical data from \citealtads{2001ApJ...548..296W} and concept by \citealtads{1977ApJ...217..425M}) with a density of the dust material of $\rho_\text{dust} = 2.7\, \rm g\,cm^{-3}$. Each SPH particle represents a dust grain size distribution following the powerlaw $\text{d}\, n(a_\text{dust})/\text{d}\, a_\text{dust} \propto {a_\text{dust}}^{-3.5}$ \citepads{1969JGR....74.2531D}. The selected range of grain sizes was $a_\text{dust} \in \left[0.005\, \mu \textup{m}, 100 \, \mu \textup{m}\right]$.
Dust of this composition and size distribution is frequently found in young stellar objects (e.g.,\citealtads{2003ApJ...588..373W}; \citealtads{2012A&A...543A..81M}; \citealtads{Graefe13}). The absolute number density of dust grains was scaled according to the total mass of dust in the disk.\par

{\em Central binary star:}
Both components of the binary system were considered as radiation sources. Additional possible heating sources were neglected (e.g., viscous heating). The stars were assumed to be blackbody radiators that are characterized by their effective temperature and luminosity. The primary component is a Herbig Ae type star with a luminosity of $L = 43 \, \rm L_\odot$, an effective temperature of $T = 9500\, \rm K$, and a stellar mass of $M = 2.5\, \rm M_\odot$. The binary mass ratio $q$ defines the properties of the secondary component as listed in Table \ref{tab:stars}.
\begin{table}[h]
\caption{Properties of the secondary component for the selection of binary mass ratios from Sect. \ref{sec:sph} \label{tab:stars}}
\centering
\begin{tabular}{lrrr}
\hline \hline
$q$ &  $L$ & $T_\textup{eff}$ & $M$ \\
  &  {[L$_\odot$]}  &	{[K]}	& {[$\textup{M}_\odot$]} \\
\hline
1 & 43 & 9500 & 2.5\\
0.5& 0.95 & 4000 & 1.25\\
0.1& 0.7 & 4000 & 0.25\\
0.01& 0.003 & 2700 & 0.025\\
\hline
\end{tabular} 
\end{table}

\begin{figure}[t]
    \resizebox{\hsize}{!}{\includegraphics{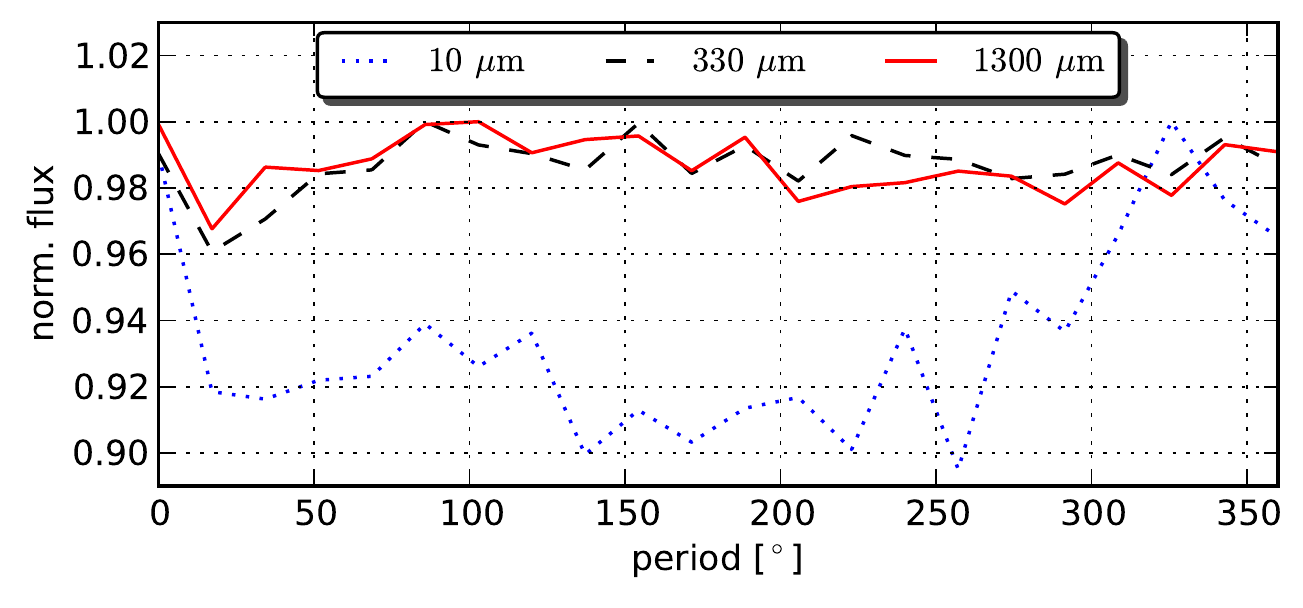}}
\caption{Fluctuations in the thermal re-emission flux of the disk during one period of the binary. The binary mass ratio is $q = 0.5$, the eccentricity $e = 0.1$. Different line types and colors represent different wavelengths.}
\label{fig:rotation1}
\end{figure}
\paragraph{Fluctuating thermal disk structure:}
The asymmetries resulting from the binary-disk interaction have affect the ability of the stars to illuminate the disk and therefore the temperature profile. The disk temperature profile is characterized by high variations in vertical, azimuthal, and radial direction. 
We discuss these variations on the example of a disk model with the parameters $q= 0.1$, $i= 30^\circ$, and $e = 0.0$. 
Figure \ref{fig:vertical_disk_temp} shows vertical cuts through the temperature distribution of this disk along the zone of the lowest and highest disk deformation ($\varphi = 131^\circ$ and $\varphi = 40^\circ$, respectively). At a distance of $5\, \rm AU$ from the primary, the ratio of the surface temperatures along cuts is $2.6$. The higher temperature along $\varphi = 40^\circ$ is the result of a better illumination of this area through direct stellar light. The higher temperature leads to a higher thermal dust re-emission in this region.

\begin{figure}[t]
    \resizebox{\hsize}{!}{\includegraphics{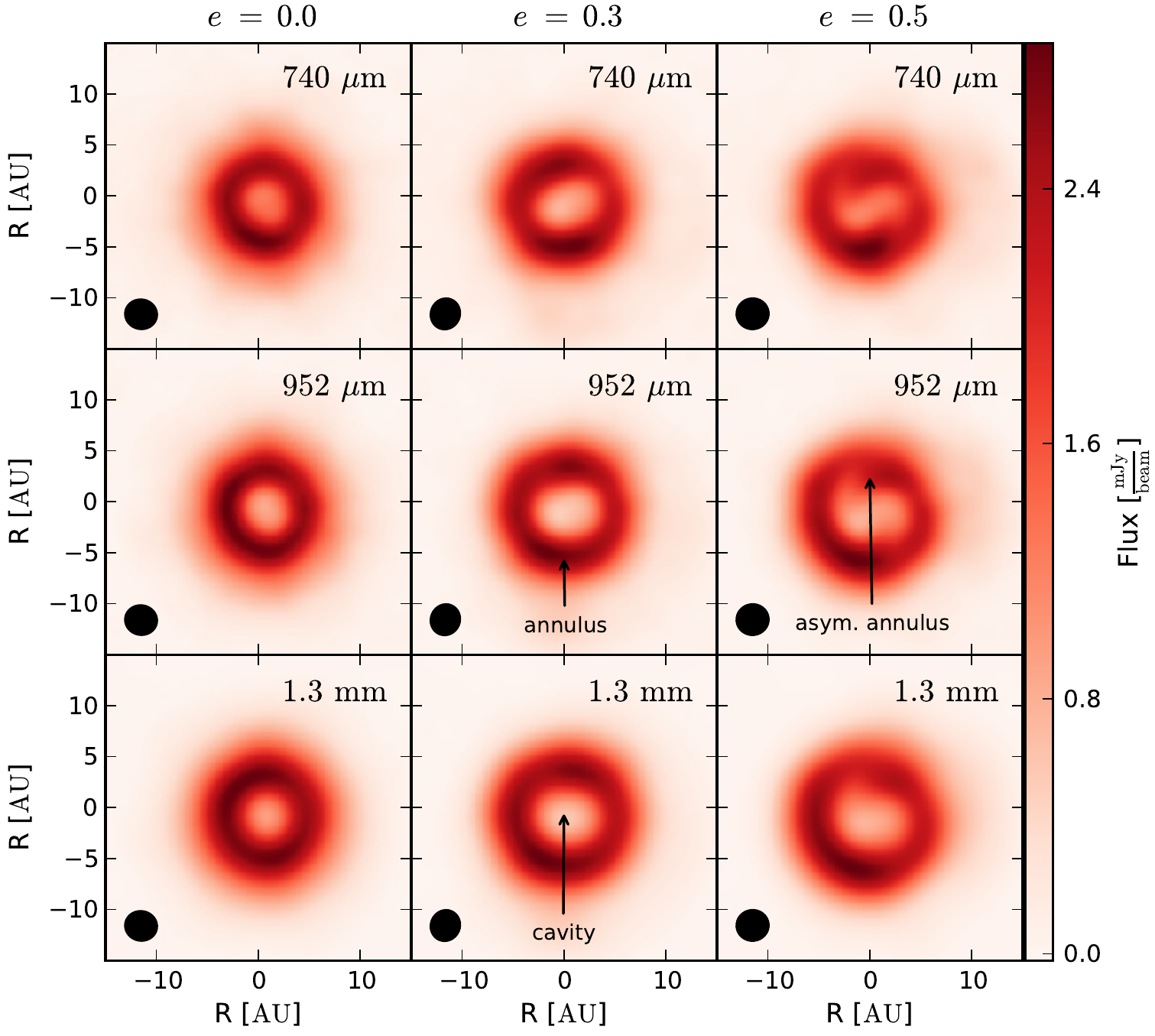}}
\caption{Simulated ALMA maps at different wavelengths for eccentric binary orbits. The binary mass ratio is $q=0.1$. The synthesized beam size of ALMA is plotted in the lower left corner of each image. The disks are identified by an annulus.}
\label{fig:ecc_ALMA_q01}
\end{figure}
%
\paragraph{Reemission light-curve:}
\label{sec:radia-res}
As a result of the variable illumination described above, not only the density structure but also the temperature profile and thus the resulting reemission flux of the disks is time-dependent.
We discuss this effect on the example of a disk model with a binary mass ratio of $q = 0.5$ and an eccentricity of $e = 0.1$.
For wavelengths  of $10\, \rm \mu m$, $330\, \rm \mu m$, and $1300\, \rm \mu m$, the thermal re-emission light-curve of this disk is presented in Fig.\ref{fig:rotation1} 
in 21 snapshots during one orbit of the central binary ($1.7\, \rm yr$). 
Although the rotation of the binary is periodic, there is no periodicity in the light curves at any wavelength on the scale of the orbital period of the binary. 
This behavior is connected to the temporal evolution of the density waves and spiral arms in the disk. 
During the orbital motion of the central stars, the structure of a surrounding disk changes, for example spiral arms are created and vanish again. Locally, this results in a variation of dust mass per dust temperature interval and leads to a temporal dependence of the re-emission flux of the disk. The effect of variations of the dust temperature on the resulting reemission flux is strongest at the shortest wavelengths.
In agreement with this, we find that the amplitude of the fluctuation decreases with increasing wavelength. 
More specficially, the variations in the (sub)millimeter range are in the order of $2\, \%$, but reach up to $10\, \%$ at mid-infrared wavelengths. 
Furthermore, the magnitude and temporal evolution of the reemission radiation depend on the observing wavelength, tracing regions with characteristic temperatures in the disk. 
Equivalent fluctuations in the scattered-light regime are observed in detail for example in the system of HH30 \citepads{2000ApJ...542L..21W,2008MNRAS.387.1313T}.

\begin{figure}[t]
    \resizebox{\hsize}{!}{\includegraphics{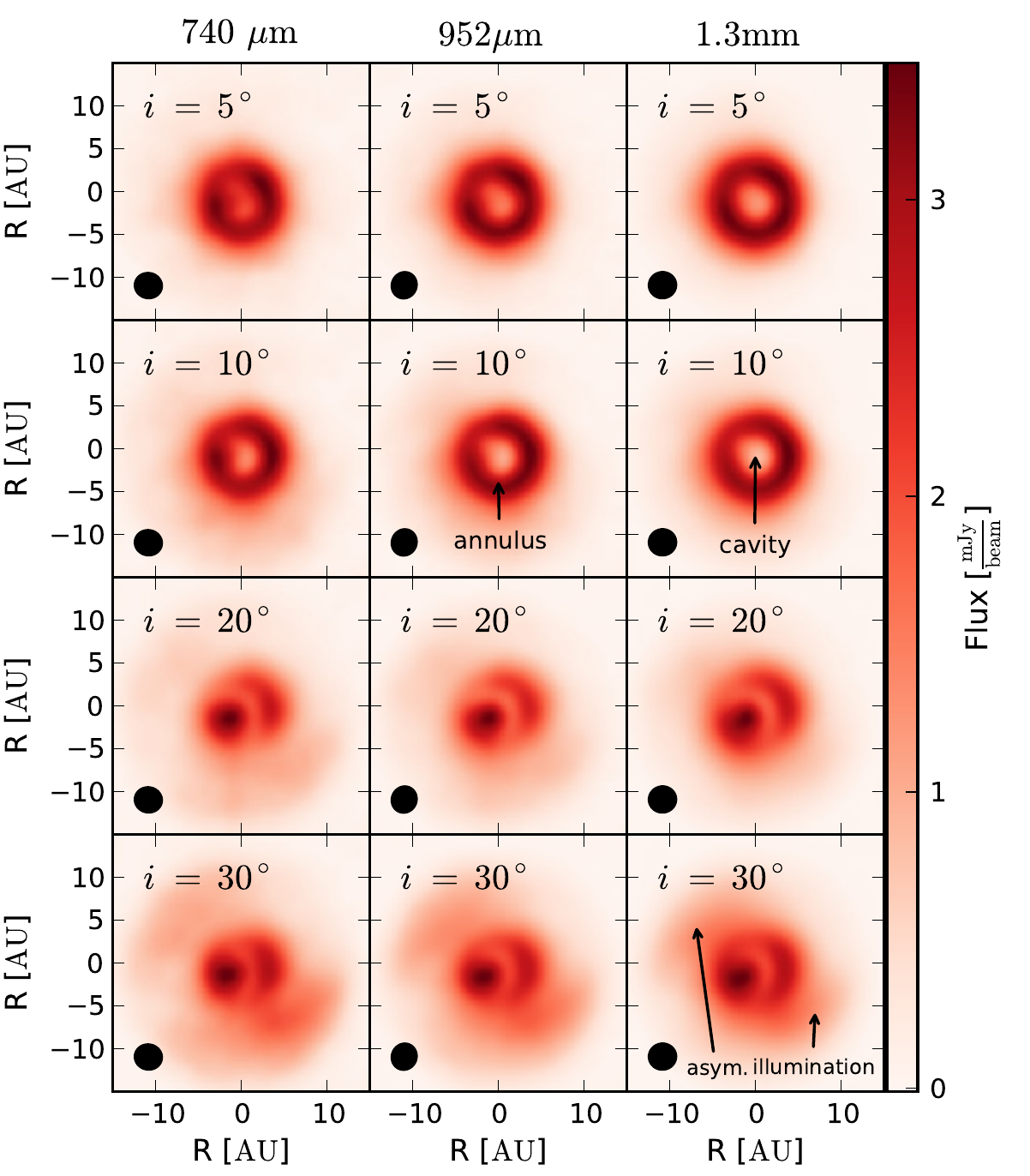}}
\caption{Simulated ALMA observations for systems with binary orbits that are inclined with respect to the midplane of the initial disk. The binary mass ratio is $q=0.1$. The synthesized beam size of ALMA is plotted in the lower left corner of each image. The detection of ring-like structures strongly depends on the disk inclination.}
\label{fig:incl_ALMA1}
\end{figure}

\section{ALMA observing setup}
\label{sec:ALMA}

\begin{figure*}[t]
\includegraphics*[width = 8.5cm]{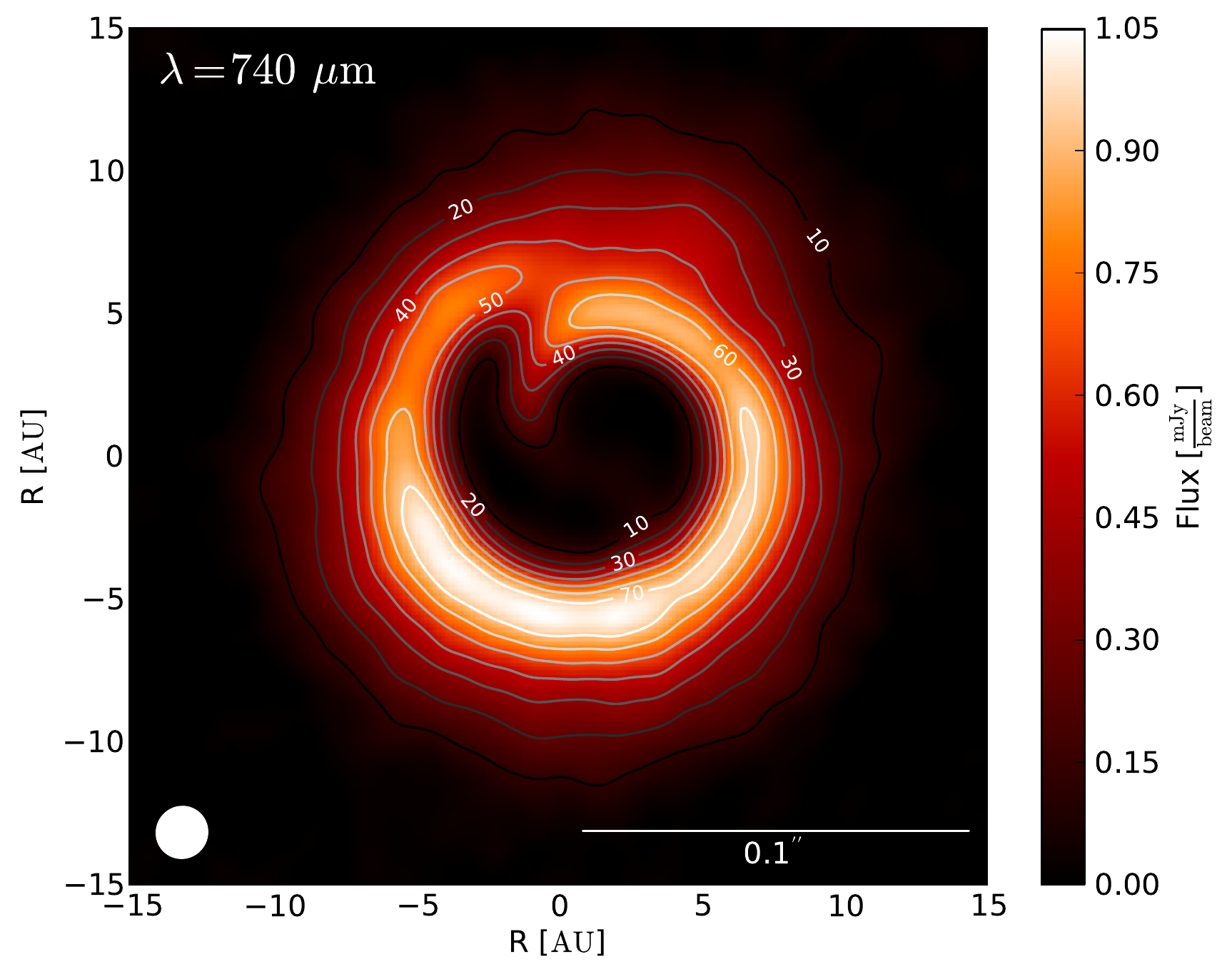}
\includegraphics*[width = 8.5cm]{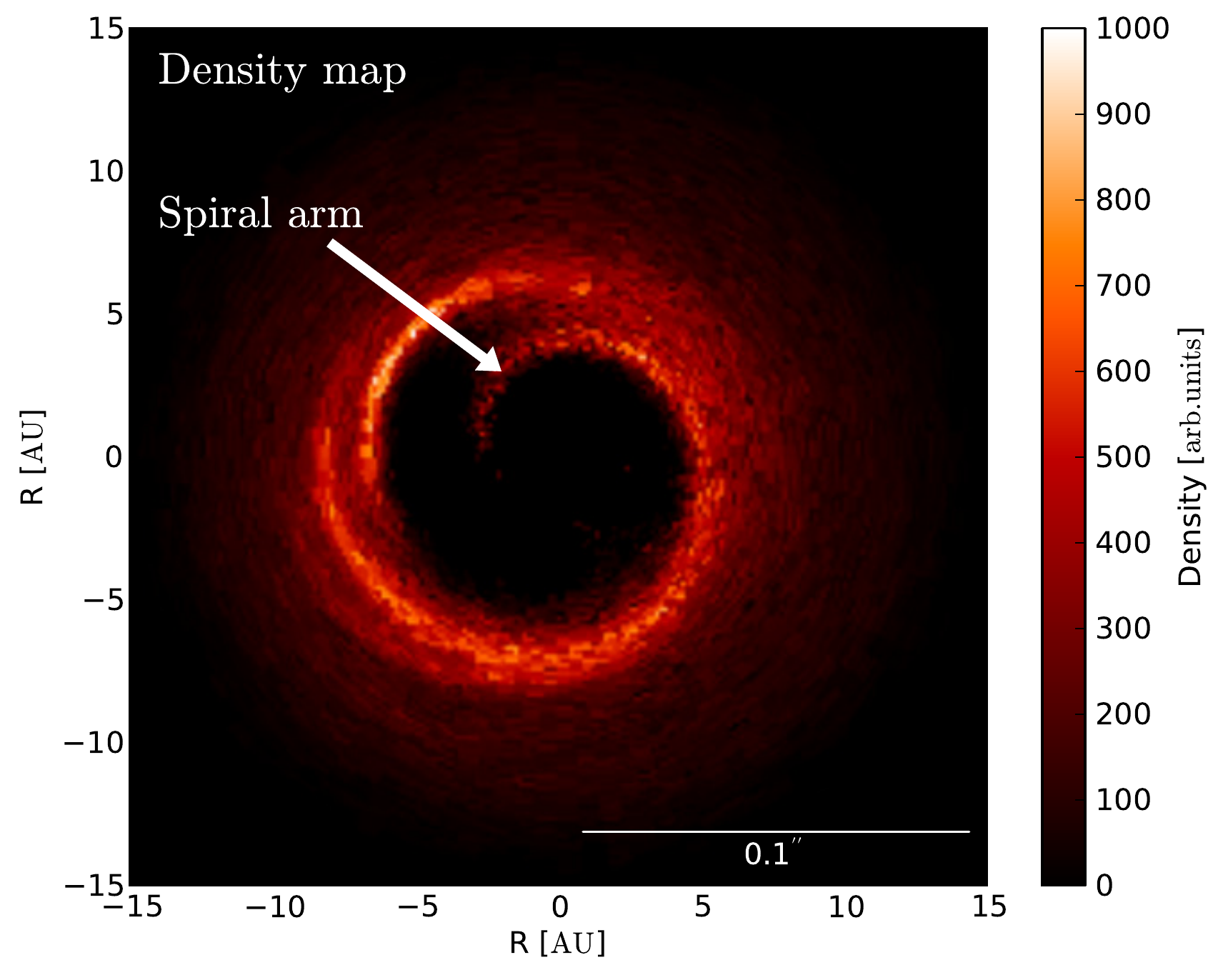}\\
\caption{Detectability of the spiral arm features in a circumbinary disk (model parameters: $q= 0.5$, $i= 0^\circ$, and $e = 0.3$). Left: $\lambda = 740\, \rm \mu m$, $t_\text{exp} = 2 \times 5\, \rm h$, max. baseline: $\approx 15\, \rm km$, contour lines for signal-to-noise (S/N) level plotted in the figure. Right: corresponding midplane density map in arbitrary units.}
\label{fig:res-spiral}
\end{figure*}
Since the wavelength, object position, and exposure time are already fixed, the specific configuration of the ALMA antennas is left as the only free parameter. We aim at resolving at least structures on an angular scale of $0.07\arcsec$ ($\approx 10\, \rm AU$ at a distance of $140\, \rm pc$) and thus only consider ALMA configurations tat fulfill this requirement (i.e., at $1.0\, \rm  mm$ a max. baseline $\geq 3500\, \rm m$). Additionally, the S/N of a synthetic observation map has to be greater than five for further use. Accordingly, the S/N is determined from the background signal in the simulated observations. If there is more than one possibility of an observation that fulfills these criteria, we chose the simulated observation with the highest resolution for each wavelength. Consequently, the beam size is of similar size at all wavelengths.

We simulated ALMA observations with the CASA 4.2 simulator \citepads{2012ASPC..461..849P} for the (arbitrary) reference position of the Butterfly Star (IRAS 04302+2247, $\alpha =$ 04h33m, $\delta = $ +22$^\circ$53$\arcmin$, J2000) in the Taurus-Auriga star-forming complex at a distance of $140\, \rm pc$. Thermal and phase noise awere considered. The detailed numerical setup for a wavelength between $740\, \rm \mu m$ and $1.3\, \rm mm$ is explained in Appendix \ref{a:casa}. Shorter wavelengths were not considered because earlier studies indicated that the very extended ALMA configurations will not allow achieving a sufficiently high signal-to-noise ratio \citepads{2013A&A...549A..97R}.

\subsection{Results}
\label{sec:ALMA-res}

\subsubsection{Circumbinary disks in a face-on orientation}
\label{sec:ALMA-res-face}
For all considered orbital elements ($i < 20^\circ$) of the binary system and binary mass ratios $q > 0.001$, ALMA is able to detect the inner cavity of the disks (see Figs. \ref{fig:ecc_ALMA_q01} and \ref{fig:incl_ALMA1}). In particular for low eccentricities and inclinations the disks appear as bright annuli in the simulated ALMA maps. The size of the inner cavity is identical for all considered wavelengths. An increase of the eccentricity of the binary orbits leads to asymmetries in the brightness distribution of the rings (Fig. \ref{fig:ecc_ALMA_q01}). Spiral arms are not traced with this ALMA setup. The behavior of disk with binaries on inclined orbits ($i< 20^\circ$) is similar. Therefore it is not possible to distinguish between binaries on moderately eccentric or inclined orbits. Only for higher inclinations ($i \ge 20^\circ$) is the structure of the disks in the simulated ALMA maps different from all other cases. As already discussed in Sect., \ref{sec:radia}, the vertically warped structured of the disk leads to zones of higher temperatures -- and therefore higher re-emission -- in some of the outer parts of the disks. ALMA is able to trace this feature (see Fig. \ref{fig:incl_ALMA1}).
For binary systems with a binary mass ratio $q = 0.001$ the simulated ALMA maps do not show any evidence of a perturbation.

\paragraph{Detection of binary-induced spiral arms:} 
In the density maps (Figs. \ref{fig:density_ec} and \ref{fig:density_inc}) spiral arms are the most prominent feature in the disk. Therefore, we investigated whether it is possible to trace these binary-induced spiral arms with ALMA.
For an example, we selected a disk model with the parameters $q= 0.5$, $i= 0^\circ$, and $e = 0.3$. It is characterized by a prominent spiral arm feature that is conserved in the re-emission maps. For the required resolution we focused on a wavelength of $740\, \rm \mu m$ and explored the most extended configurations of the ALMA array ($\# 28$ in the CASA array configuration files) with a longest baseline of about $15\, \rm km$. We calculated the observing time with the ALMA Sensitivity Calculator, resulting in $2 \times 5\, \rm h$ to achieve a root mean square noise of $0.013\, \rm mJy$.\par
The result is plotted and compared to the corresponding density  map in Fig. \ref{fig:res-spiral}. We added contour lines to the figure that indicate the S/N levels.
With ALMA it is possible to trace a spiral arm in our circumbinary disk models of $2 \times 3 \, \rm AU^2$ with the S/N areas that are shown in Fig. \ref{fig:res-spiral}.

\paragraph{Discussion:}
In general, one should be aware of tracing from structures in (simulated) ALMA observations back to the real origin of these structures. Indeed, ALMA is able to resolve and trace the spiral arms that are prominent in our disk models, but from the ALMA observation alone one will not be able to conclude on the binary nature of the central objects.

\begin{figure*}[t]
\includegraphics*[width = 9cm]{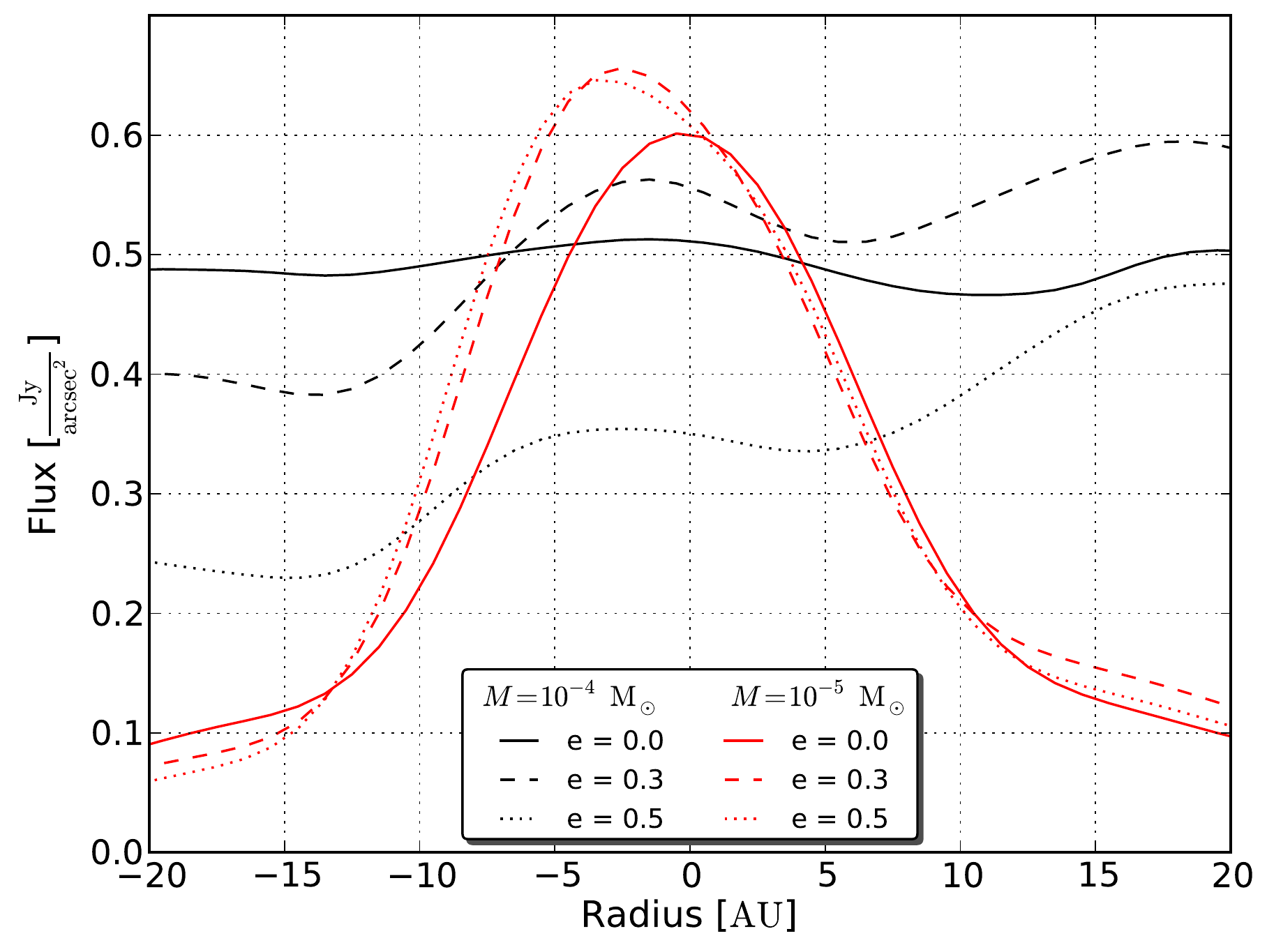}
\includegraphics*[width = 9cm]{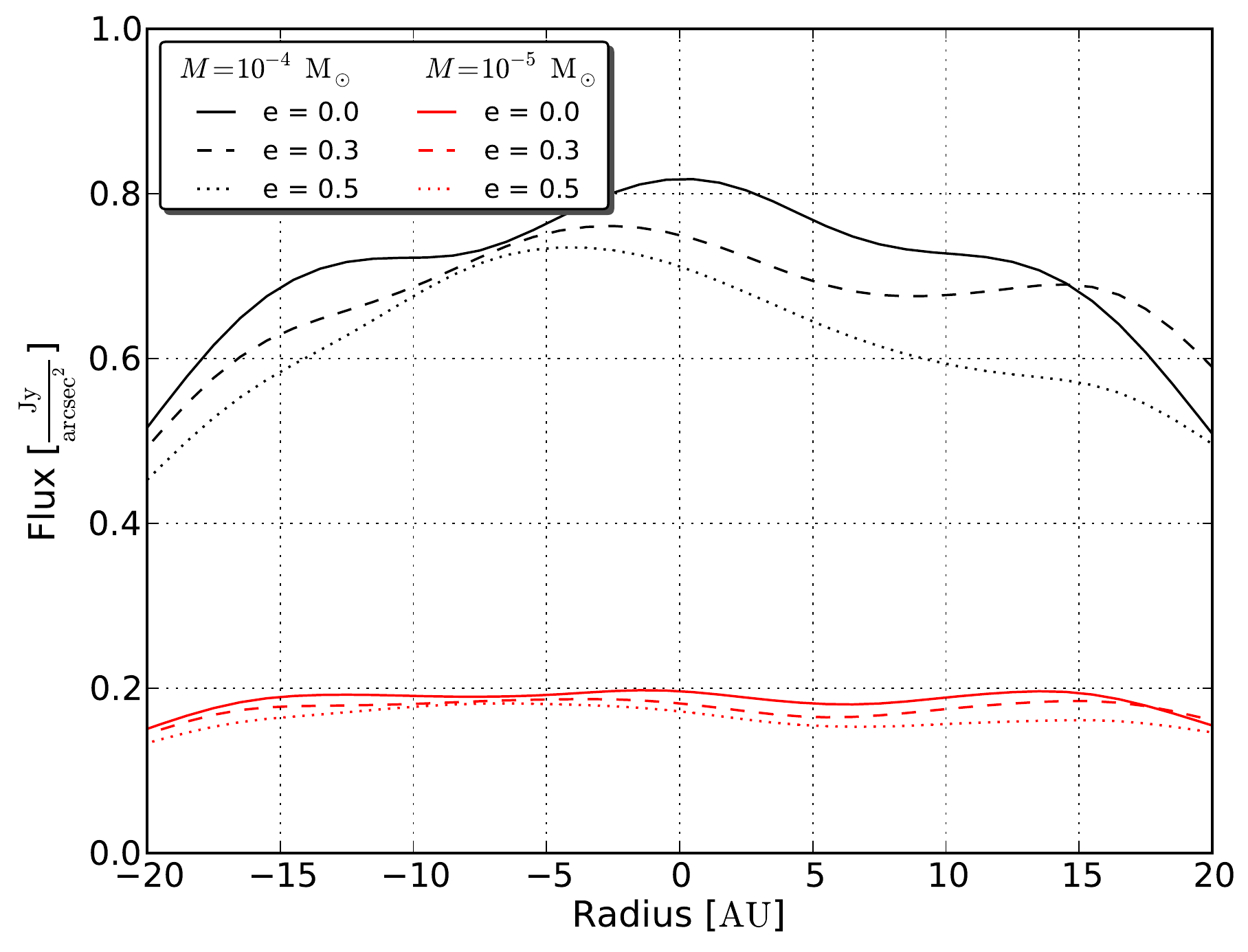}
\caption{Edge-on brightness profile of disks around an eccentric binary. Observing wavelength: $1.3\, \rm mm$. The longest baseline of the ALMA configuration is $\approx 8\, \rm km$. Left: binary mass ratio $q =0.1$; right: binary mass ratio $q =0.01$. Red: $M_\text{dust} =  10^{-5} \, \rm M_\odot$, black: $M_\text{dust} =  10^{-4} \, \rm M_\odot$.}
\label{fig:ec_ALMA_edge}
\end{figure*}
\begin{figure*}[t]
\includegraphics*[width = 9cm]{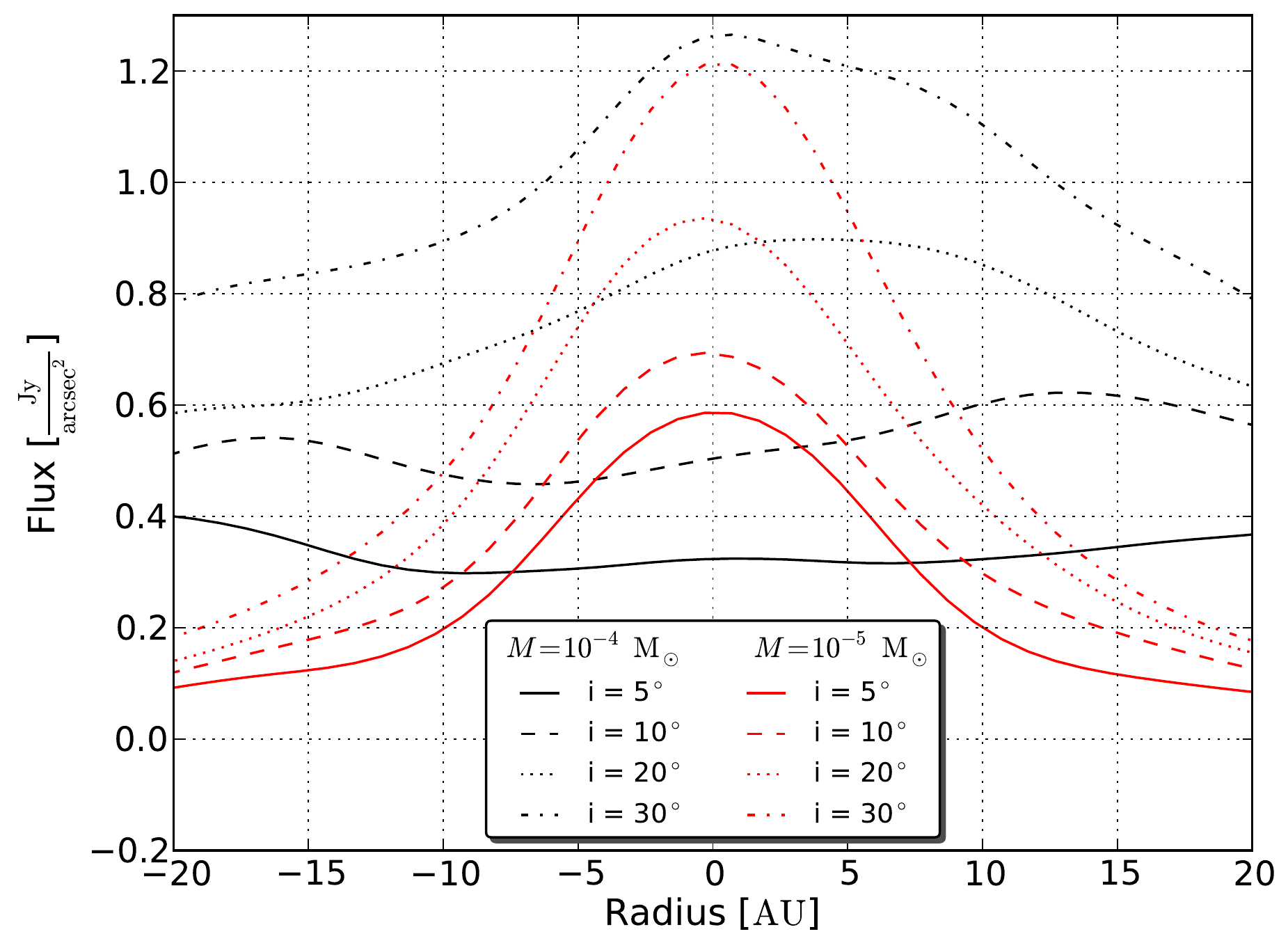}
\includegraphics*[width = 9cm]{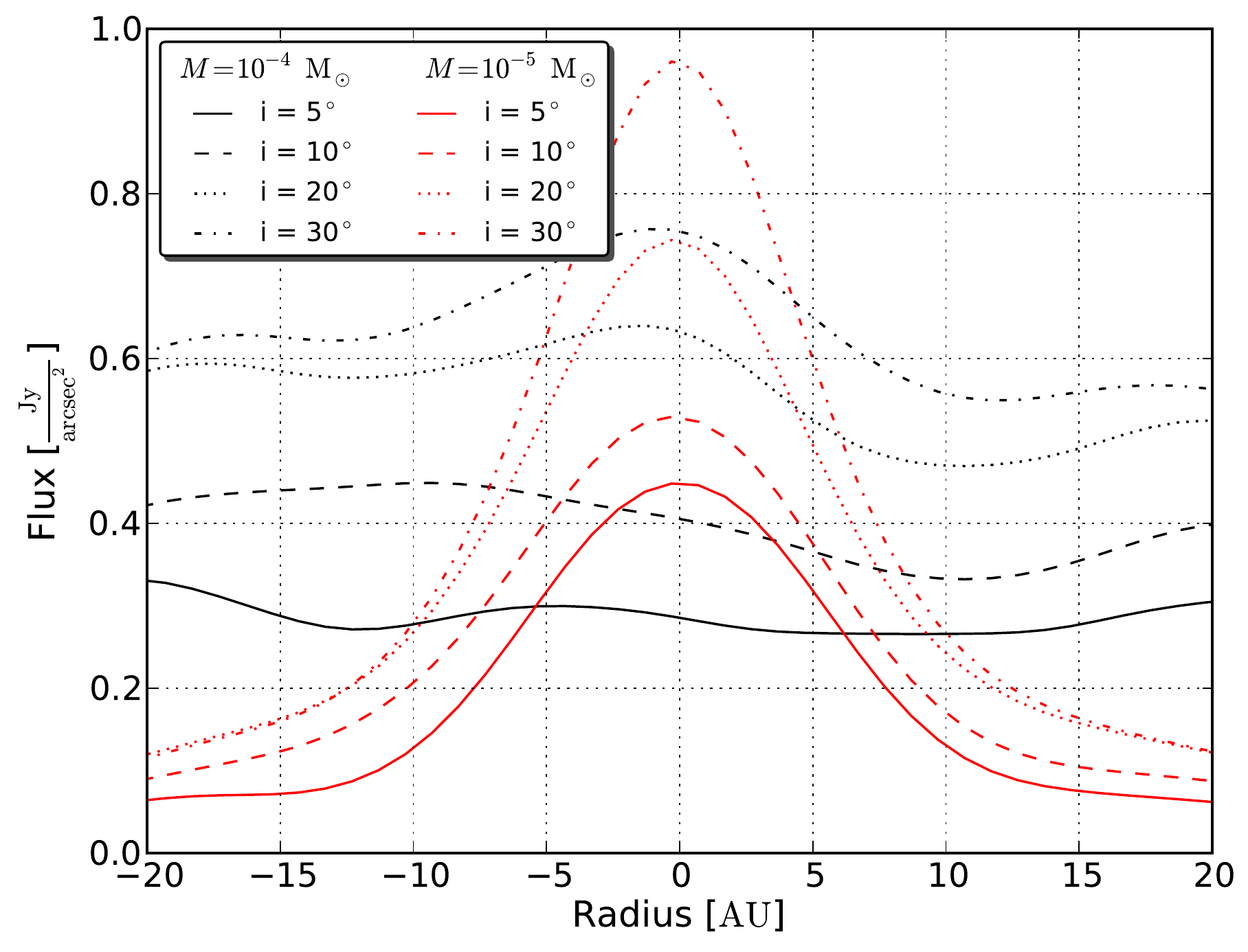}
\caption{Midplane brightness profile of binaries seen edge-on that are inclined to the initial disk midplane. Wavelength: $1.3\, \rm mm$. The longest baseline of the ALMA configuration is $\approx 8\, \rm km$. Left: binary mass ratio $q =0.1$; right: binary mass ratio $q =0.01$. Red: $M_\text{dust} =  10^{-5} \, \rm M_\odot$, black: $M_\text{dust} =  10^{-4} \, \rm M_\odot$.}
\label{fig:inc_ALMA_edge}
\end{figure*}
\subsubsection{Circumbinary disks in an edge-on orientation}
\label{sec:ALMA-res-edge}

\paragraph{Required ALMA array configuration and data analysis:}
Now we changed the disk orientation to edge-on and determined whether it is possible to trace features of the binary perturbation of the disk in the simulated ALMA maps. We focused on a wavelength of $1.3\, \rm mm$ and an ALMA configuration ($\# 22$) with a longest baseline of $\approx 8\, \rm km$ of the ALMA array. This combination results in a synthesized beam of about the same size as the vertical extension of our disk models and maps with the highest possible S/N. From the simulated observations we extracted a horizontal cut through the disk and plot the resulting flux against the distance from the center (radial brightness profile). To avoid mistakes in the interpretation of the data that are possibly based on the fitting process of the analytically described outer disk onto the numerically simulated inner disk (see Sect. \ref{sec:sph}), we only focused the radial brightness profile on a range of $20\, \rm AU$ around the center of the disk. Additionally, the radial brightness profile is influenced by the orientation of the disk to the observer. We investigated the range of inclinations of the objects from $85^\circ$ to $95^\circ$ and found the following results to be valid from $88^\circ$ to $92^\circ$.
The considered dust masses of the disk are $M_\text{dust} =  10^{-5} \, \rm M_\odot$ and $M_\text{dust} =  10^{-4} \, \rm M_\odot$ (see Sect. \ref{sec:sph}).\par This mass selection marks the transiting of the disks from optically thin to optically thick along the line of sight from the observer through the edge of the disk to its center (i.e., from an averaged, effective optical depth: $\tau_{10^{-5} \, \rm M_\odot,\, 1.3\, \rm mm} = 0.65 \pm 0.04$, highest value in disk midplane $\tau_{10^{-5}\, \rm M_\odot, \text{max}} = 0.88$, to an averaged, effective optical depth: $\tau_{10^{-4} \, \rm M_\odot,\, 1.3\, \rm mm} = 4.16 \pm 0.50$, highest value in the disk midplane $\tau_{10^{-4}\, \rm M_\odot, \text{max}} = 8.79$).\par
The profile of the radial brightness distribution is a byproduct of the optical depth structure of the disk. It is thereby directly linked to the disk density profile. In our perfect edge-on case an asymmetry in the radial brightness profile therefore is the result of asymmetries in the density distribution of the disk, which is here induced by the binary.
\par
We find that the influences of inclined and eccentric binary orbits are partially different, therefore we explicitly differentiate between these two cases in the following.

\paragraph{Eccentric binary orbits:}
For binary mass ratios $q = 0.1$ and $0.01$ the radial brightness profiles for eccentric binary orbits are plotted in Fig. \ref{fig:ec_ALMA_edge}.
For $q=0.1$ our disk models have inner cavities that become noncircular with increasing eccentricity of the binary orbit (see Fig. \ref{fig:density_ec}). The form of the cavity and the spiral arms, which are density waves that propagate through the disk, create a radial brightness profile that is not symmetric to the disk center. In particular, the brightness profiles of the models with  $e = 0.0$ (circular cavity) and of the models with $e = 0.5$ (noncircular cavity) present increasing asymmetries with increasing eccentricity of the binary orbits. The peak flux does not depend on the eccentricity of the binary orbit. Additionally, disks with $M_\text{dust} =  10^{-4} \, \rm M_\odot$ have lower peak fluxes than disks with $M_\text{dust} =  10^{-5} \, \rm M_\odot$. This is explained by an optical depth effect because for the more massive disk the signal from the warm inner layer of the disks is hidden by the outer parts. For disk dust masses of $M_\text{dust} =  10^{-5} \, \rm M_\odot$ the binary nature of the systems leads to an increasing offset of the flux maximum with an increasing eccentricity.\par
In the special case of systems with $q=0.01$ (Fig. \ref{fig:ec_ALMA_edge}, \textit{right}) the binary was not able to form an inner cavity. Because of this, the disks with $M_\text{dust} =  10^{-4} \, \rm M_\odot$ are already optically thin ($\tau_{10^{-4} \, \rm M_\odot,\, 1.3\, \rm mm} = 0.22 \pm 0.04$, $\tau_{10^{-4}\, \rm M_\odot, \text{max}} = 0.75$). Therefore, the local minima in the radial brightness profile do not result from an optical depth effect. Their locations can be associated with the position of a gap formed by the secondary component in the density distribution of the disk (see Fig. \ref{fig:density_ec}). Thereby, the asymmetry of the gap is due to the eccentricity of the binary, which is also be traced by the radial brightness profile. Since the disk was optically thin before, the decrease of the re-emission flux for disks with dust masses of $M_\text{dust} =  10^{-5} \, \rm M_\odot$ can be explained by the decrease of the disk mass. However, the feature in the radial brightness profile resulting from the gap in the disk density distribution is conserved. \par
Since the SPH calculations are mainly sensitive to the binary mass ratio $q$, a giant planet ($M_\text{planet} = 10\, \rm M_\text{Jup}$) around a solar-type star ($M_\star = 1 \, \rm M_\odot$) will induce similar perturbations in the disk \citepads[for a detailed study see][]{2013A&A...549A..97R}.
\par
In general, the presented results are also valid for other disk sizes and masses as long as the optical depth along the line of sight from the disk center through the edge of the disk to the observer is similar to the values shown above.

\paragraph{Inclined binary orbits:}
In particular, the brightness profile is not axis-symmetric to the rotation axis of the disk anymore (see Fig. \ref{fig:inc_ALMA_edge}).
In contrast to the case of eccentric binaries, the peak flux increases with increasing inclination of the binary orbit for both considered disk masses in our setup. The reason is the warped surface of the disks, which leads to a better irradiation of some parts of the disk (see Sects. \ref{sec:radia} and \ref{sec:ALMA-res-face}).\par

\paragraph{Discussion:}
The radial brightness profile depends on the global density distributions of the disk. Thus, large-scale perturbations, inner holes, spiral arm features, but also the total disk mass, affect the shape of the profile. In addition to disk perturbation through binary systems, cavities can form through photoevaporation or planets \citepads[e.g.,][]{2012ApJ...747..103E,2011ApJ...736...85U,2010MNRAS.407..181C}. Therefore, it is not feasible to safely trace from a radial brightness profile back to its origin.

\section{Conclusion}
\label{sec:conclusion}
For a simple circumbinary disk model, we investigated the feasibility of tracing characteristic density structures in circumbinary disks that result from the binary-disk interaction. For close binary systems, we analyzed the impact of the binary mass ratio, its eccentricity, and its inclination on the appearance of the global disk structure. For this purpose, we simulated the 3D density structure of the circumbinary disk (SPH simulations), calculated the corresponding dust temperature structure and reemission images (radiate transfer simulations), and derived simulated ALMA observations (CASA 4.2).
Our main results are as follows:
\begin{itemize}
 \item The (sub)millimeter reemission brightness distribution is characteristic for given binary mass ratios $q$. However, for $q \leq 0.001$ the effect of the binary nature, the resulting variable gravitational potential, and thus its impact, is too weak to create structures that are prominent enough to be traced with ALMA.
 \item Binaries on highly inclined orbits create a warped density profile that leads to illumination conditions and temperature profiles of the disk with large variations in azimuthal direction. Within the parameter space considered in this study, the surface temperature can locally be increased by a factor of $2.6$ as compared to unperturbed disks.
 \item The time-dependent illumination leads to temporal flux variations in the thermal re-emission of the disk on the order of a few percent. The fluctuations are directly linked to the appearance of density waves and spiral arms in the disk.
 \item The inner cavity (e.g., $q > 0.001$) can be traced at wavelengths $\ge 740\, \rm \mu m$ with ALMA for all disks in our setup and assuming standard observing conditions.
 \item The simulated ALMA observations show the inner edge of the disks as a bright ring with local asymmetries.
 \item Only at $740\, \rm \mu m$ is it possible to trace the binary-created spiral arms of a few $\rm AU$ in size in the disk.
 \item For disks in a perfect edge-on orientation, the simulated ALMA maps indicate asymmetries in the disk structure.
\end{itemize}
We emphasize that we considered the binary-circumbinary disk interaction as the sole source of the studied disk structures. Other mechanisms, such as planet-disk interaction (\citealtads{2005ApJ...619.1114W}; \citealtads{2012A&A...547A..58G}; \citealtads{2013A&A...549A..97R}), magnetic fields (e.g., \citealtads{2014arXiv1411.2736F}), self-gravitation structures (\citealtads{2010MNRAS.407..181C}), vortices (\citealtads{2002ApJ...578L..79W}), and photoevaporation (\citealtads{2012ApJ...747..103E}) may result in density waves and cavities as well, but were not considered in this study. The feasibility of distinguishing between spiral density waves with different physical origins, for example through observations of different accompanying effects are suggested for future investigations.

\begin{acknowledgements}
T. Demidova and V. Grinin are supported by the RFBR grant 15-02-09191. T. Demidova was supported by the DFG grant WO 857/4-2. J.P. Ruge is supported by the DFG grant WO 857/10-1. The project was supported by the DFG grant WO 857/12-1. We thank N. Sotnikova for SPH calculation code and discussion of the models and P. Scicluna for language corrections. The main fpart of the plots was made with the Python matplotlib \citep{Hunter:2007}.
\end{acknowledgements}

\bibliography{biblio}

\appendix

\section{CASA setup}
\label{a:casa}
ALMA is a (sub)mm interferometer in the Atacama desert for observations in the wavelength range between $0.3\, \rm mm$ and $9.3\, \rm mm$. We considered the full ALMA array, which can be arranged in different configurations \citepads{2004AdSpR..34..555B}.
CASA offers a mighty toolkit for data reduction, analysis, and simulation for ALMA and other (sub)mm interferometers. We predicted the observations with the CASA 4.2 simulator \citepads{2012ASPC..461..849P}.
To simulate an ALMA observation, CASA calculates the visibilities in the u-v-plane as sampled by the array configuration, adds noise, and finally reconstructs the image with the CLEAN algorithm. The observing wavelengths listed in Table \ref{tbl:wavel} were selected to suit the atmospheric windows in the (sub)mm range. The bandwidth for the (continuum) ALMA observations is $8\, \rm GHz$, the selected exposure time is two hours.

\begin{table}[htbp]
\caption{Selected wavelengths for the simulated ALMA observations. \label{tbl:wavel}}
\begin{center}
\begin{tabular}{ccccc}
\hline\hline
ALMA band &  $\lambda$  &   $\nu$ &PWV & Octile\\
 & {[$ \rm \mu m$]} & {[$\rm GHz$]} & {[$\rm mm$]}   &   \\
\hline
8 & 740 & 405 & 0.658 $\pm$ 0.1 & 2nd \\
7 & 952 & 315 & 0.913  $\pm$ 0.1 & 3rd \\
6 & 1300 & 230 & 1.262  $\pm$ 0.15 & 4th\\
\hline
\end{tabular}
\tablefoot{PWV: precipitable water vapor}
\end{center}
\end{table}
We considered the influence of thermal noise by precipitable water vapor (PWV) in the atmosphere using the wavelength-dependent values listed in Tab. \ref{tbl:wavel} and the \textit{tsys-atm} option in the CASA simobserve task. Additionally, we included the influence of atmospheric phase noise in our calculations. 
We assumed a phase-screen blowing with a constant velocity above the ALMA array with the PWV values and deviations listed in Table \ref{tbl:wavel}. This is numerically done in CASA with the \textit{sm.settrop} task in screen mode.

\end{document}